%% file: main.tex
\newcommand{\liqdata}{liquid data}
\newcommand{\Liqdata}{Liquid data}

\newcommand{\ourprotocol}{BitRipple protocol}

\newcommand{\oursoftware}{BitRipple software}
\newcommand{\Oursoftware}{BitRipple software}

\newcommand{\oursolution}{BitRipple solution}

\documentclass[sigconf, nohyperxmp]{acmart}

\AtBeginDocument{%
  \providecommand\BibTeX{{%
    \normalfont B\kern-0.5em{\scshape i\kern-0.25em b}\kern-0.8em\TeX}}}

\begin{document}

\title{Enabling immersive experiences  \\ in challenging network conditions}


\newcommand{\nsfgrantnote}{This work was supported in part by the National Science Foundation under 
Grant No.~2212574. Any opinions, findings, and conclusions or recommendations 
expressed in this material are those of the authors and do not necessarily 
reflect the views of the National Science Foundation.}

\author{Pooja Aggarwal}
\authornote{\nsfgrantnote}
\affiliation{%
  \institution{BitRipple, Inc.}
  \city{Berkeley}
  \country{USA}
}
\email{pooja@bitripple.com}

\author{Michael Luby}
\authornotemark[1]   
\orcid{0000-0002-6239-8072}
\affiliation{%
  \institution{BitRipple, Inc.}
    \city{Berkeley}
\country{USA}}
\email{luby@bitripple.com}

\author{Lorenz Minder}
\authornotemark[1]   
\affiliation{%
  \institution{BitRipple, Inc.}
    \city{Berkeley}
\country{USA}
}
\email{lorenz@bitripple.com}

\renewcommand{\shortauthors}{Aggarwal, Luby, Minder}

\begin{abstract}
  Immersive experiences, such as remote collaboration and augmented and virtual reality, require delivery of large volumes of data with consistent ultra-low latency across wireless networks in fluctuating network conditions.  We describe the high-level design behind a data delivery solution that meets these requirements and provide synthetic simulations and test results running in network conditions based on real-world measurements demonstrating the efficacy of the solution.
\end{abstract}





\maketitle

\section{Introduction and Background}
Immersive experiences, such as remote collaboration and augmented and virtual reality, require delivery of large volumes of data with consistent ultra-low latency across wireless networks in fluctuating network conditions.

One common approach to data delivery, employed by real-time protocols such as webRTC~\cite{rfc8834}, \cite{rfc8854}, \cite{rfc8627}, and QUIC~\cite{rfc9002}, is to retransmit packets when there is packet loss.  Retransmission-based solutions might not work well when reliable ultra-low latency delivery of data is required.  For example, the delivery latency can fluctuate dramatically depending on the actual packet loss: When a packet is lost en route, the receiver identifies the lost packet and sends a packet identifier of the lost packet in feedback to the sender so that the sender knows to retransmit the packet, which adds at least a round-trip time (RTT) to the delivery time of the packet.  If the packet is lost en route again when the sender retransmits the packet, then the receiver again identifies the lost packet and sends the packet identifier in feedback to the sender so that the sender can retransmit the packet again.  Depending on the amount of packet loss between the sender and receiver, this process may repeat several times before the packet is delivered to the receiver, causing several RTTs of data delivery latency.  

It can be shown that data delivery latency variability is inherent with retransmission-based solutions, thus it is difficult for retransmission-based solution to achieve consistently reliable ultra-low latency data delivery.
One approach to that problem is to send the original streaming data in packets and accept, or hide as well as possible, displayed artifacts when not all of the streaming data is available at the receiver due to packet loss.  However, as streaming video data becomes more highly compressed and higher quality, if even a tiny portion of the streaming video data is not available at the receiver then the resulting artifacts when it is displayed are typically unacceptable.

Another approach is to use a forward error correction erasure code at a fixed code rate to protect against packet loss up to a specified amount, e.g. for webRTC, see~\cite{rfc8834}, \cite{rfc8854}, \cite{rfc8627}, \cite{rfc5109}.  This method needlessly consumes bandwidth when there is little or no packet loss, whereas data is not reliably delivered when the amount packet loss exceeds the specified amount of packet loss protection over the specified duration of time.  This approach can be combined with protocols to retransmit lost source packets when the loss exceeds the forward error correction protection~\cite{rfc8627}, although then the deliver latency reverts to that of a retransmission-based protocol.

An example of a work that is similar in some ways to ours is~\cite{RFEC22}, where reinforcement learning and the WebRTC feedback protocol is used to determine the rate at which the original source multimedia stream should be sent and determine the amount of FEC protection to send to provide packet loss protection for the stream.  We concentrate on determining the amount of FEC protection to send, and in future work we will focus on determining the rate at which the original source stream should be sent.

Our goal with respect to FEC protection seems to be a bit different than that of~\cite{RFEC22}. Our protocols are designed to proactively send slightly more FEC protection than needed so as to minimize the worst case delivery latency, whereas~\cite{RFEC22} concentrates more on providing a good average deliver latency and minimizing the amount of bandwidth used for FEC protection by relying more on retransmission.  For example, in the results shown in Table 2 of~\cite{RFEC22}, the worst case delivery latency of the R-FEC protocols introduced in~\cite{RFEC22} are comparable to or worse than those of previous solutions.  

We describe a data delivery solution that satisfies immersive experience requirements. The essential idea is similar to that described in Liquid Data Networking~\cite{LDN20}, i.e., erasure codes are integrated into the foundation of the solution for delivering real-time data streams with consistently minimal latency and close to minimal bandwidth usage in varying network conditions.

We provide a high level protocol design,  synthetic simulation results, a high level description of a full software implementation together with performance results, and network simulation results based on employing the software in network conditions chosen based on real-world data.

\section{High Level Protocol Design}

Based on the RaptorQ fountain code \cite{rfc6330}, \cite{Raptorbook}, all data sent over the network is RaptorQ encoded data, which we hereafter refer to as {\bf \liqdata}. \Liqdata\ is data generated from a data object that is {\bf expandable} and {\bf interchangeable}. 
\begin{itemize}
    \item {\bf Expandable} means that any amount of \liqdata\ can be generated on the fly from a data block, which means that any amount of network packet loss can be overcome.
    \item {\bf Interchangeable} means that portions of \liqdata\ carried in packets are interchangeably valuable in recovery of the data block, and thus it doesn’t matter what portions of \liqdata\ are lost, it only matters that the total amount of \liqdata\ that arrives at the receiver is at least the data block size.  
\end{itemize}

\begin{figure}[h]
  \centering
  \includegraphics[width=\linewidth]{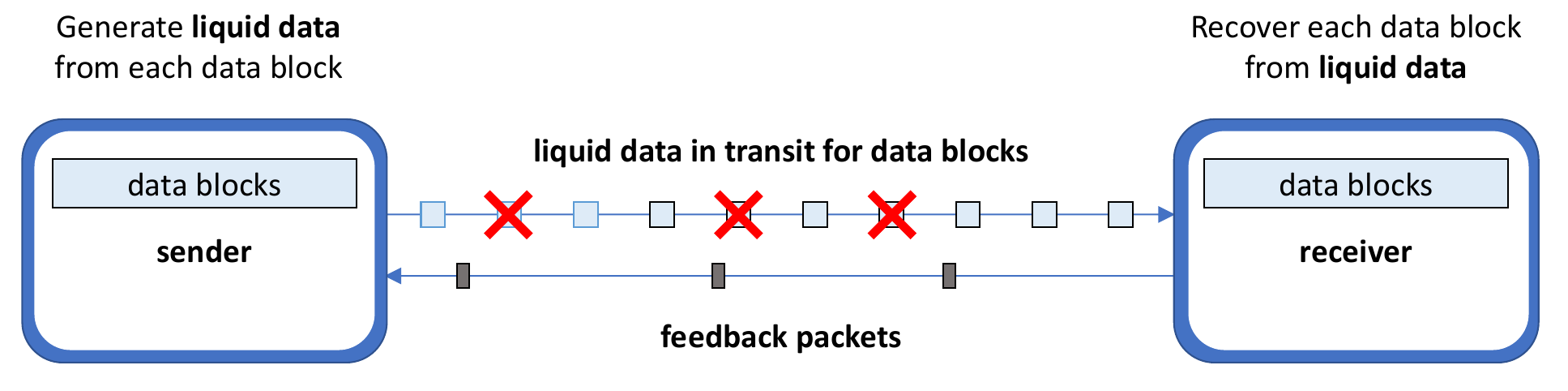}
  \caption{Basic design of the \ourprotocol}
  \Description{}
  \label{protocoldesign}
\end{figure}

The high level design of the overall protocol, hereafter referred to as the {\bf \ourprotocol}, is shown in Figure~\ref{protocoldesign}.  As each data block arrives to the sender, 
the sender automatically generates \liqdata\ from the block and transmits the \liqdata\ to the receiver. The sender sends the \liqdata\ in UDP packets with a header that includes a sequence number that is global across all transmitted packets (similar to QUIC~\cite{rfc9002}), a block identifier, an identifier of what \liqdata\ for the block is contained in the packet, and the block size. 

As indicated by the red Xs shown in Figure~\ref{protocoldesign}, packets may be lost in transmission between the sender and receiver.

The receiver recovers each data block as soon as the amount of \liqdata\ that has arrived for the block is equal to the block size.  In addition, the receiver continually sends feedback packets to the sender including information such as the largest global sequence number that has been received in a packet, the total number of packets containing \liqdata\ that have been received, and the amount of \liqdata\ that has been received for each active block,  where a block is active at the receiver if at least some \liqdata\ has been received for the block but not enough \liqdata\ has yet been received to recover the block.  

Packets are never retransmitted.  Instead, based on feedback from the receiver to the sender, the sender generates and sends an initial amount of \liqdata\ for each data block suitable for current network conditions. If the feedback indicates that network conditions have worsened while delivering a block, then the sender generates and sends additional \liqdata\ for the block. The sender continually adjusts the amount of \liqdata\ generated and transmitted for each data block based on feedback from the receiver, using algorithms that ensure reliable delivery of data blocks while at the same time minimizing delivery latency and bandwidth usage.  

The objectives are to send an amount of \liqdata\ for each block so that both {\bf delivery latency} and 
{\bf bandwidth usage} are minimized.
\begin{itemize}
    \item{\bf Delivery latency minimization:} The amount of \liqdata\ that arrives at the receiver is at least the block size, which allows recovery of the block without packet retransmissions.
    \item{\bf Bandwidth usage minimization:} The amount of \liqdata\ that arrives at the receiver is not much more than the block size, which minimizes bandwidth usage.  
\end{itemize}
\textbf{The sender proactively sends an amount of \liqdata\ for a data block so that the amount of \liqdata\ expected to arrive at the receiver for that block, given current network conditions, is slightly more than the data block size.} 

The algorithms that the sender uses to decide how much \liqdata\ to send for each data block are based on estimates of the current packet loss rate, the variance in the packet loss rate, and how much \liqdata\ has already been reported to have been received in feedback for the data block.  
The sender protocols ensure that close to the minimal amount of \liqdata\ required to recover a data block arrives at the receiver, thus ensuring efficient network bandwidth usage.

\section{Synthetic simulation results}

In this section we present a Python simulation of how the \ourprotocol\ operates in a simplified simulation of a network. For comparison, we also provide a simulation of the optimal retrans\-mission-based protocol operating in the same network conditions.

This simplified simulation is not fully representative of how a real network operates, e.g., in the simplified simulation we assume a given number of "packet transmission slots" between the arrival of each frame, whereas in a real network there are no "packet transmission slots".  We present this simplified simulation since it was instrumental to the design of \ourprotocol, and since it provides a method to directly compare \ourprotocol\ to retransmission-based protocols.

The retransmission-based protocol we simulate is optimal in the sense that whenever a packet is lost in transmission, the receiver is provided the knowledge that the packet is lost at the point in time when the packet would have arrived if it had not been lost, the receiver immediately sends a retransmission request to the sender, and the sender immediately responds and retransmits the packet as soon as the request arrives.  Thus, the delivery latency for data blocks for the optimal retransmission-based protocol is a lower bound on the delivery latency for any protocol based on retransmission.  

The simulation is for a 150 Mbps video stream at 30 fps. Each frame is a data block in this simulation, where consecutive frames become available at the sender at 33.3 ms intervals of time.  Each frame is of equal size, i.e., 5 Mbits in size, or 625 KB. Data for the frames is sent in packets with 10,000 bit payloads, i.e., 1.25 KB payloads, and thus the source data for a frame fits into 500 packets.

There is a 16.65 ms delay on the path from the sender to the receiver and also a 16.65 ms delay from the receiver to the sender, and thus the overall RTT is 33.3 ms. 

There are 800 equally-spaced packet transmission slots available between frames. The ratio of the number of packet slots to the number of packets per frame implicitly puts an upper bound on the total available bandwidth in the simulation, in this case the available bandwidth is 1.6 times the video source bandwidth (800/500), i.e., the available bandwidth is 240 Mbps.  As packet loss is introduced in the simulation, possible goodput decreases below that of the available bandwidth.

\begin{figure}[h]
  \centering
  \includegraphics[width=\linewidth]{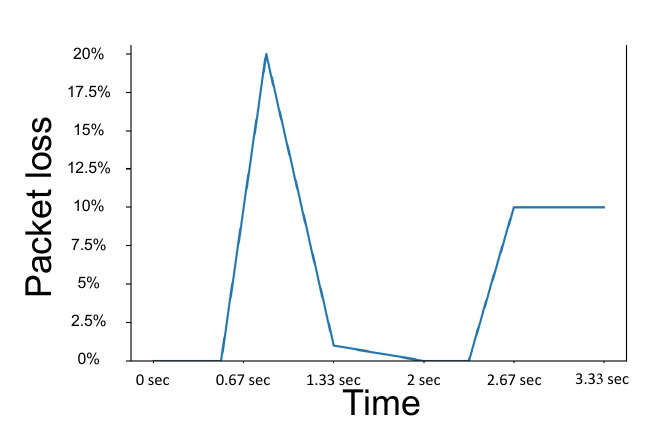}
  \caption{Varying network packet loss for a 150 Mbps video stream at 30 fps}
  \Description{}
  \label{plchart}
\end{figure}

\begin{figure}[h]
  \centering
  \includegraphics[width=\linewidth]{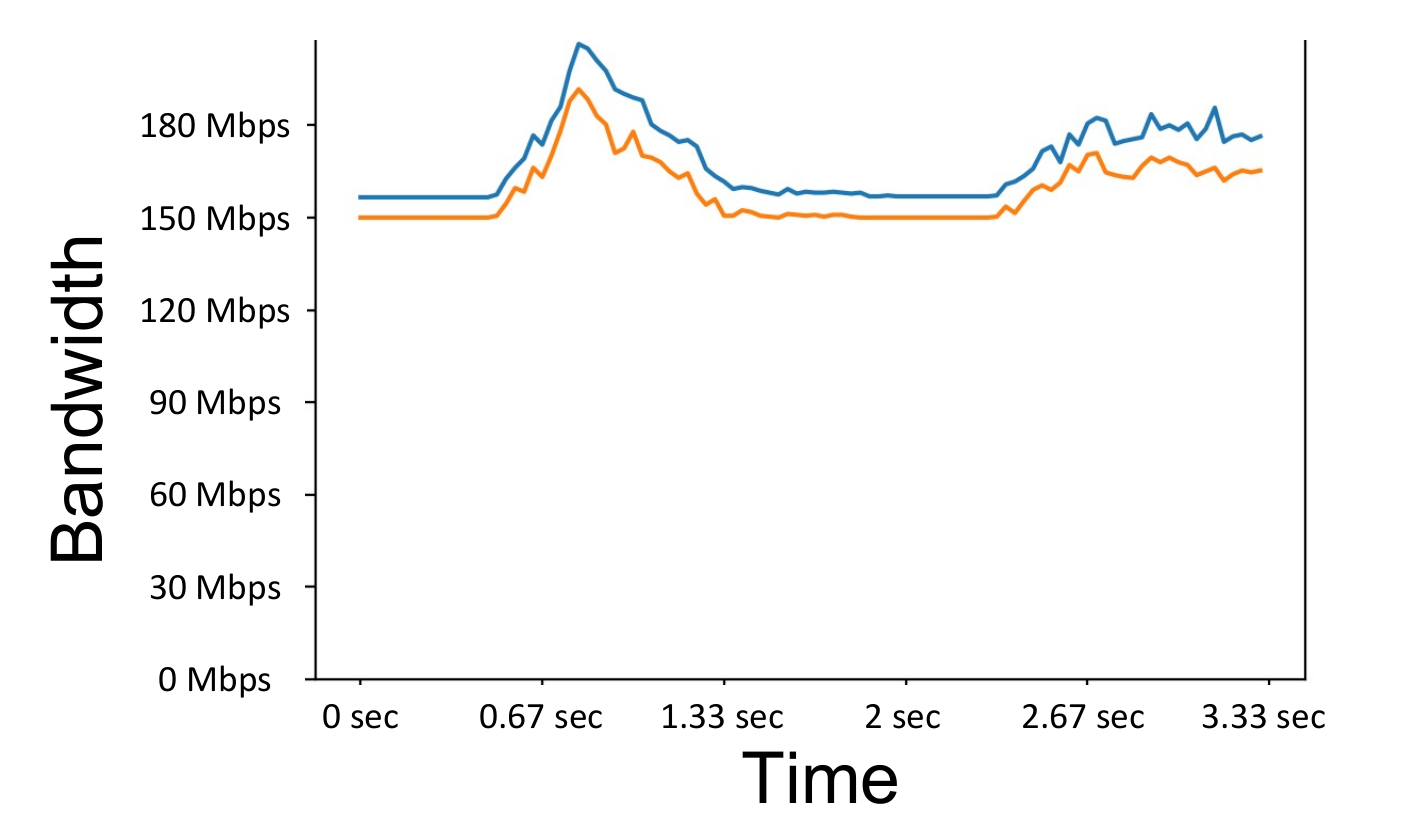}
  \caption{Transmission bandwidth for a 150 Mbps video stream at 30 fps}
  \Description{}
  \label{bwchart}
\end{figure}

\begin{figure}[h]
  \centering
  \includegraphics[width=\linewidth]{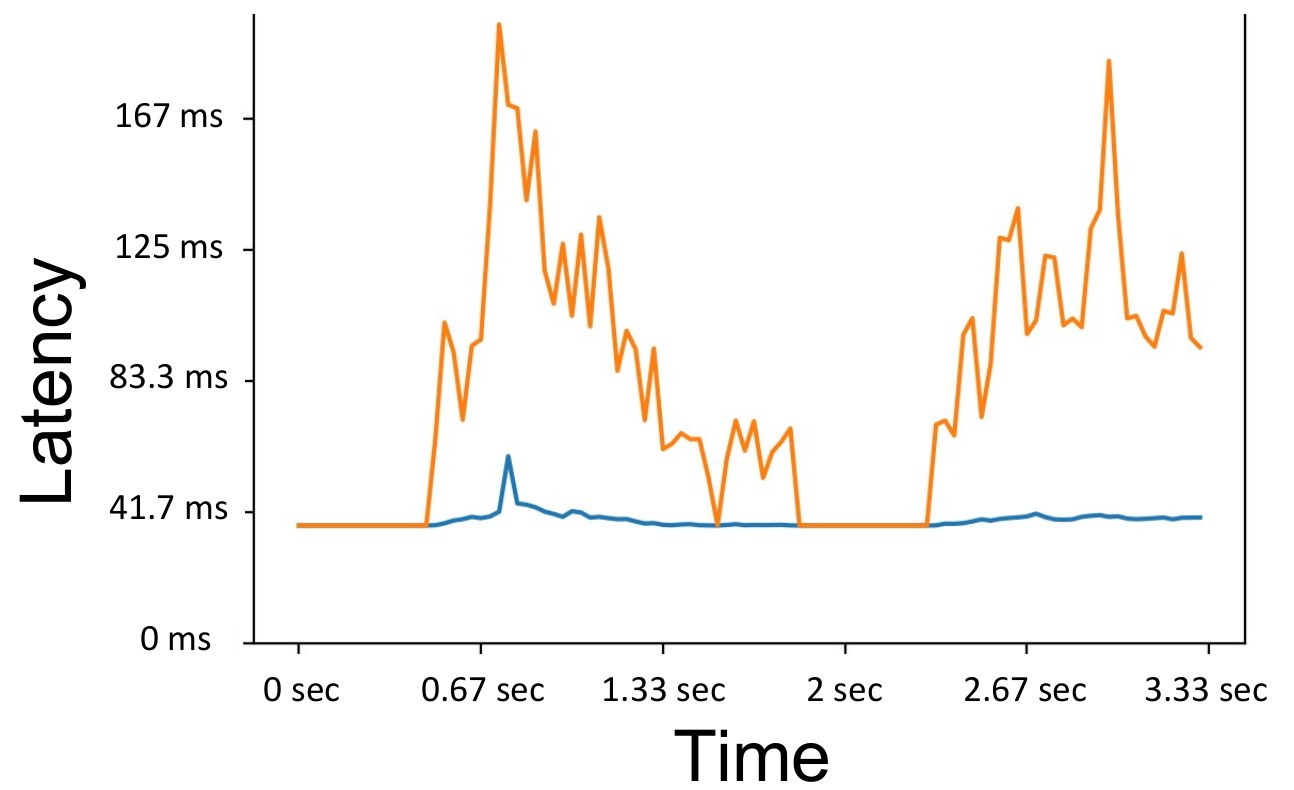}
  \caption{Delivery latency for a 150 Mbps video stream at 30 fps}
  \Description{}
  \label{dlchart}
\end{figure}

Figure~\ref{plchart} shows how the network conditions vary over time during the simulation, in terms of increases and decreases in the packet loss.  Packets in packet time slots are dropped randomly at each point in time at the rate indicated on the Y-axis.
There is no packet loss at the beginning of the simulation, and then there is a spike of packet loss that starts at 0.5 seconds, peaks at 0.83 seconds, decreases quickly until 1.33 seconds, decreases more slowly, etc. Both the \ourprotocol\ and the retransmission-based protocol are run against the same packet loss trace in the graphs shown in Figures~\ref{bwchart} and~\ref{dlchart}.

Figure~\ref{bwchart} shows the transmission bandwidth usage to reliably deliver the frames of video over time:
\begin{itemize}
    \item The {\bf orange line} in Figure~\ref{bwchart} indicates the transmission bandwidth usage for the optimal retransmission-based protocol under the network conditions shown in Figure~\ref{plchart}.  This is also a lower bound on the transmission bandwidth usage for any protocol that reliably delivers the frames under these network conditions.
    \item The {\bf blue line} in Figure~\ref{bwchart} indicates the transmission bandwidth usage for the \ourprotocol\ under the network conditions shown in Figure~\ref{plchart}.  It can be seen that the transmission bandwidth usage is above the optimal, but is always in a tight band that in this example is at most around 5\% more than the optimal transmission bandwidth usage.
\end{itemize}

Figure~\ref{dlchart} shows the frame delivery latency as frames are delivered over time.  The frame delivery latency is the time between when a frame is available for transmission at the sender until the time the frame has been recovered at the receiver:
\begin{itemize}
    \item The {\bf orange line} in Figure~\ref{dlchart} indicates the frame delivery latency for the optimal retransmission-based protocol under the network conditions shown in Figure~\ref{plchart}.  Since these results are for the optimal retransmission-based protocol, these results are also a lower bound on the delivery latency for any retransmission-based protocol that reliably delivers the frames under these network conditions (e.g., most implementations based on webRTC or QUIC).
    \item The {\bf blue line} in Figure~\ref{dlchart} indicates the frame delivery latency for the \ourprotocol\ under the network conditions shown in Figure~\ref{plchart}.  It can be seen that the delivery latency is consistently close to the minimum possible even as the network conditions vary.
\end{itemize}

\section{Software Design and Performance} 

\begin{figure}[h]
  \centering
  \includegraphics[width=\linewidth]{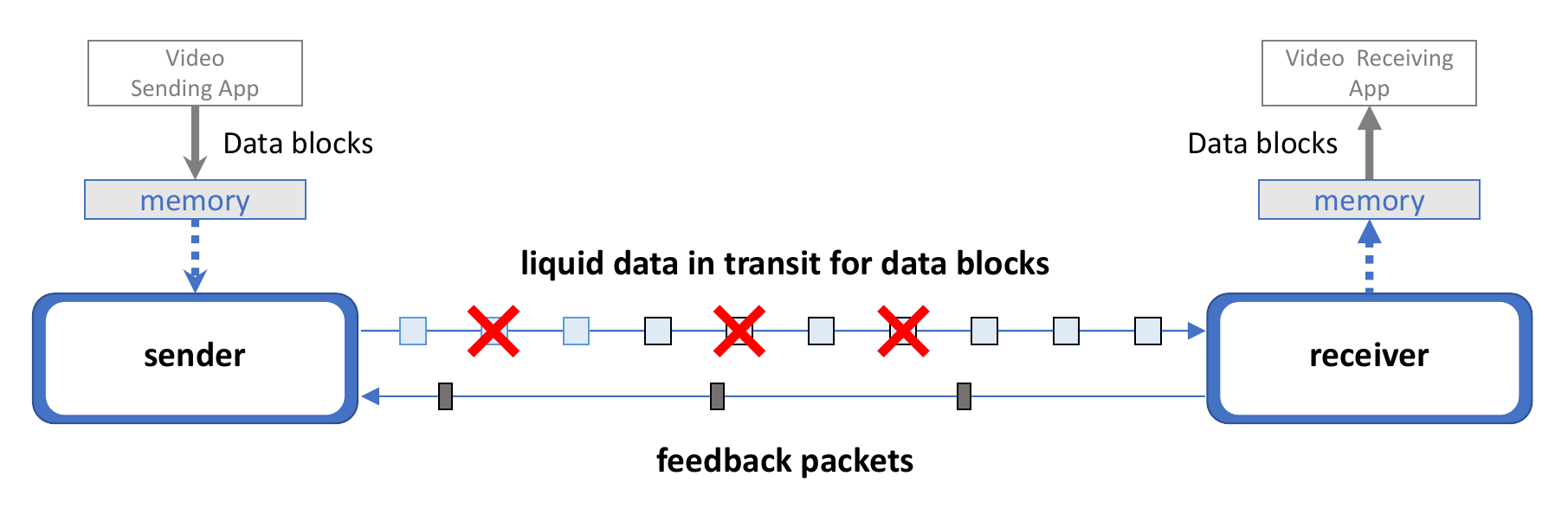}
  \caption{\Oursoftware\ interfaces}
  \Description{}
  \label{softwaredesign}
\end{figure}

We have developed a software implementation of the \ourprotocol, which we hereafter refer to as the {\bf \oursoftware}.  The interfaces to the sender and receiver software are shown in Figure~\ref{softwaredesign}.  A separate sending application, e.g., a video sending app, places each data block into a memory block and passes the pointer to the memory as well as the data block size to the sender.  The sender generates packet payload sized portions of \liqdata\ from the data block based on RaptorQ encoding, generates the UDP packets carrying the \liqdata\ portions, and sends the packets over the network.  

The receiver is essentially the inverse of the sender.  The receiver receives UDP packets carrying \liqdata\ portions for each data block from the network, uses the received \liqdata\ portions and corresponding headers to recover the data block into a memory block based on RaptorQ decoding, and passes a pointer to the memory block as well as the data block size to a receiving application, e.g., a video receiving app.

The sender and receiver are designed to operate at high speed, and to be able to concurrently handle multiple data blocks as they arrive at the sender and are being transmitted to the receiver.  Feedback is continuously sent at a low data rate from the receiver to the sender, which is used by the sender to immediately update the amount of \liqdata\ to send for blocks on a continuous basis.

\begin{figure}[h]
  \centering
  \includegraphics[width=\linewidth]{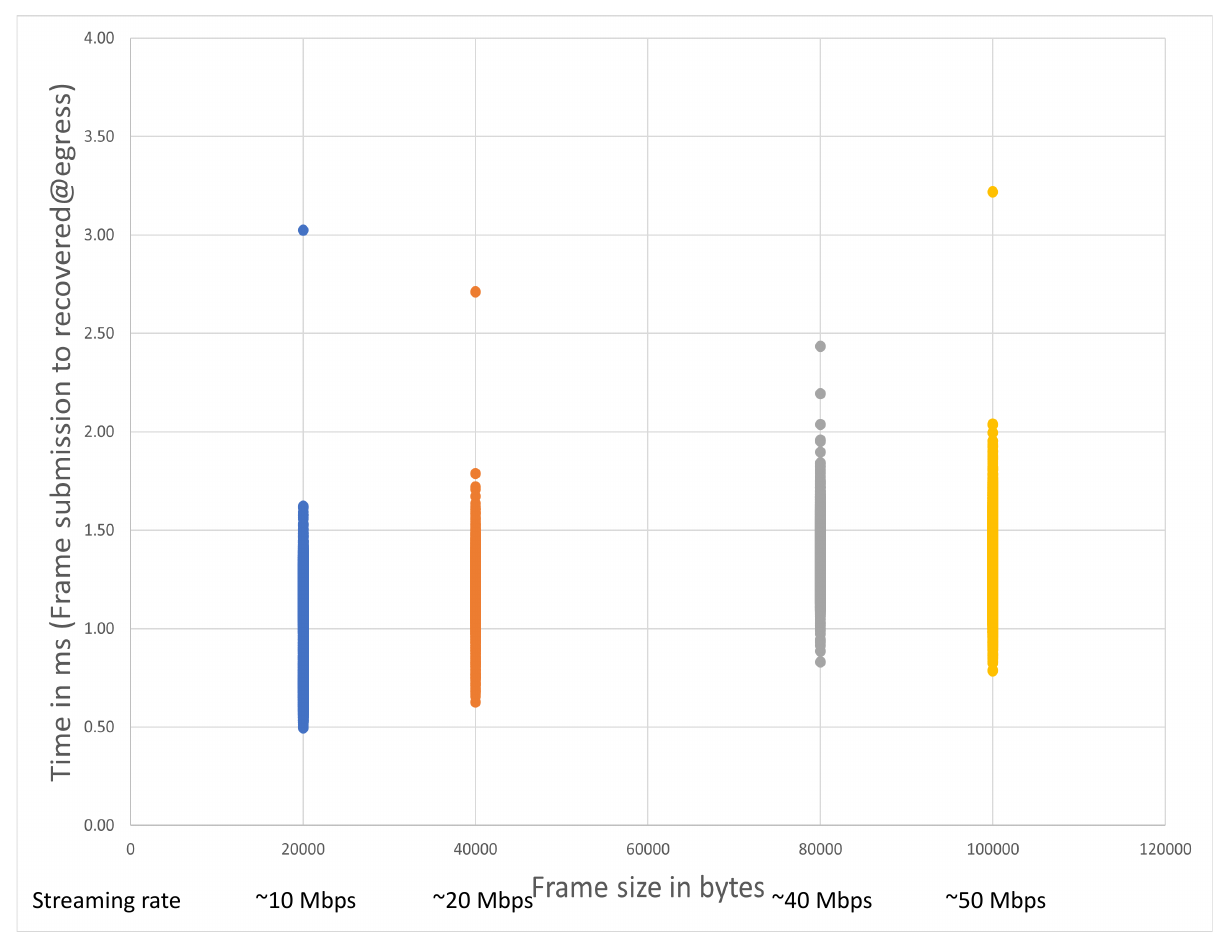}
  \caption{Overall processing latency: \oursoftware\ sender transmitting over local loopback to receiver running on same machine}
  \Description{}
  \label{softwarelatency}
\end{figure}

Figure~\ref{softwarelatency} provides some performance results that show that the overall processing latencies of the \oursoftware\ are quite small.  In these results, the sender and receiver are running on the same machine, where data blocks are provided to the sender through memory from a video sending app, the sender generates the \liqdata\ for the stream of data blocks, places the \liqdata\ into packets and sends the packets over a local loopback interface to a receiver running on the same machine.  The receiver takes received packets of \liqdata\ and recovers the stream of data blocks and provides the recovered data blocks through memory to the video receiving app. A $10\%$ packet loss is induced on the local loopback interface to ensure that the sender performs RaptorQ encoding to send data blocks and the receiver performs RaptorQ decoding to recover data blocks.  

The sender and receiver both run on the same Windows Ryzen 5700G machine. The data blocks are provided to the sender at regular 16.67ms intervals to represent a 60 fps video stream of frames. There were 10,000 equal-sized frames streamed for each of the four streaming rates. The frame size for each of the stream rates is also shown on the X-axis.  

The overall processing latencies shown in Figure~\ref{softwarelatency} are measured from the time a data block is passed through memory from the video sending app to the sender until the time when the recovered data block is passed through memory from the receiver to the video receiving app.  The time includes all the processing required for passing the data block in through memory from the video sending app to the sender, encoding, packetization, injecting packet loss, packet transmission over the local loopback interface, depacketization, decoding, and passing the data block out through memory from the receiver to the video receiving app.  At each streaming rate, there are 10,000 time results plotted.  The few outliers are suspected to be times when unrelated processes became busy on the machine.

The overall processing latency due to the combination of the sender and receiver for each data block shown in Figure~\ref{softwarelatency} is always less than about 2 ms (excluding the handful of outliers), which is much smaller than the RTT across networks of interest (see for example Section~\ref{impaired network section}).  Thus, the overall processing delays of the \oursoftware\ sender and receiver are more than made up for by the reduction in overall delivery latency compared to retransmission-based protocols. 

\section{Impaired network results}
\label{impaired network section}

\begin{figure}[h]
  \centering
  \includegraphics[width=\linewidth]{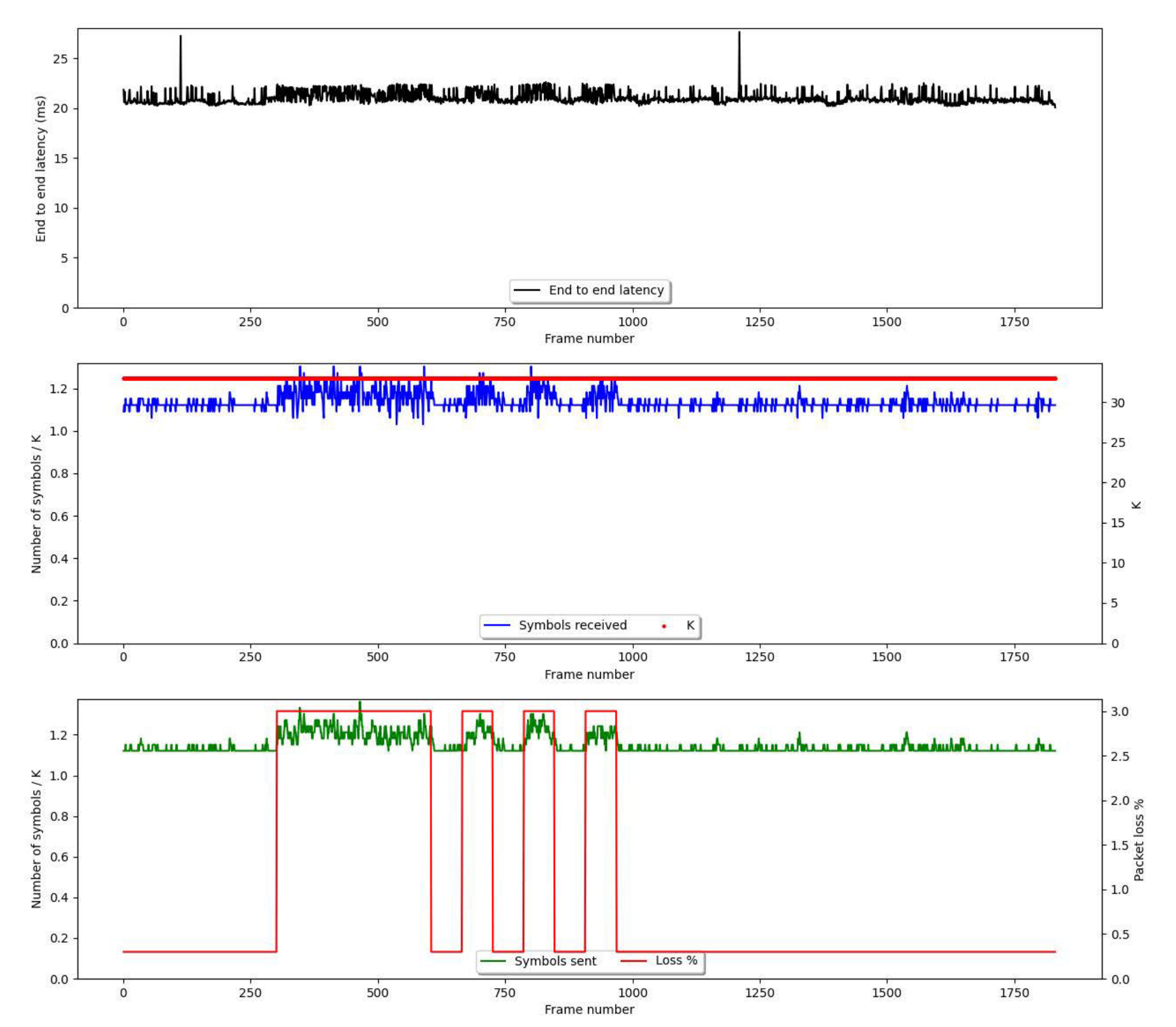}
  \caption{9.6 Mbps video stream at 30 fps, 40 KB frames, 40 ms RTT, 1 minute run}
\label{Slowfig}
\end{figure}

\begin{figure}[h]
  \centering
  \includegraphics[width=\linewidth]{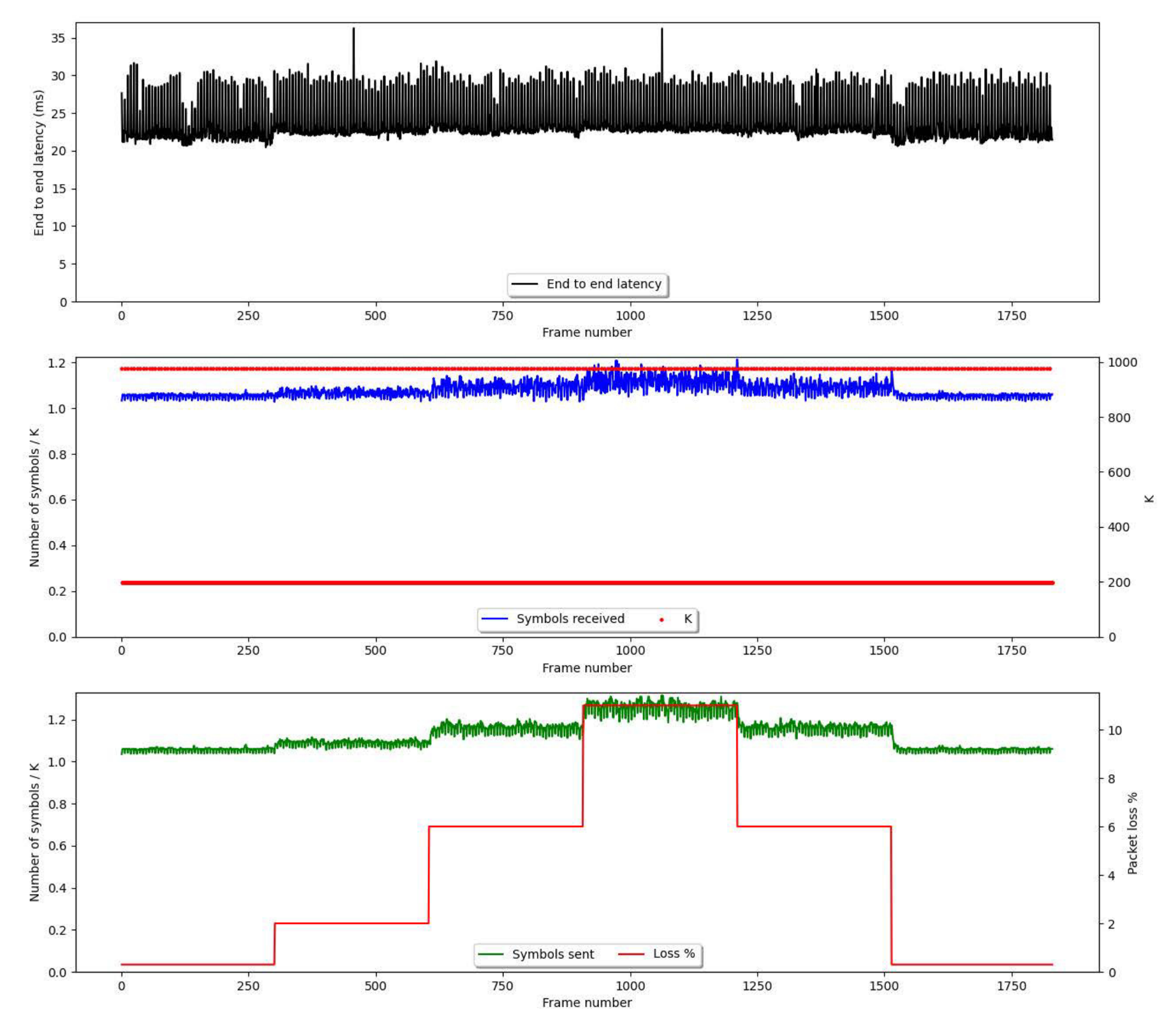}
  \caption{100 Mbps video stream at 30 fps, mix of 250 KB and 1.25 MB frames, 40 ms RTT, 1 minute run}
\label{Fastfig}
\end{figure}

A field trial/evaluation of the \oursoftware\ with a major online cloud gaming provider guided the choice of the streaming rates, packet loss rates, and network delays used in the results shown in Figures~\ref{Slowfig} and~\ref{Fastfig}, as these streaming rates, packet loss rates, and network delays are being observed in the online cloud gaming provider's current deployment. 

Figures~\ref{Slowfig} and~\ref{Fastfig} show the \oursoftware\ running over a network between a sender running on a Linux machine to a receiver running on a different Linux machine, where netem is used to add a 20 ms delay along the network path from the sender to the receiver, and netem is also used to add a 20 ms from the receiver to the sender.  Netem is also used to induce random varying packet loss along the path from the sender to the receiver.  Each run is 60 seconds in duration, i.e., 1800 frames overall at 30 fps  are delivered from the sender to the receiver in real-time.

Figure~\ref{Slowfig} shows the results for a 9.6 Mbps video stream at 30 fps, where all frames are 40 KB in size.  The packet loss was varied in steps between $0.3\%$ and $3\%$, as shown in the red line mapped to the right Y-axis in the bottom graph of Figure~\ref{Slowfig}. 

The amount of \liqdata\ by the sender for each frame is shown by the green line mapped to the left Y-axis in the bottom graph of Figure~\ref{Slowfig}, where the amount of data is normalized so that the value 1 corresponds to the frame size.  Note that as the packet loss (red line) steps between $0.3\%$ and $3\%$, the amount of \liqdata\ sent for the frames (green line) automatically adjusts to an appropriate amount to overcome the packet loss.  

The amount of received \liqdata\ at the receiver is shown by the blue line mapped to the left Y-axis in the middle graph of Figure~\ref{Slowfig}, where the amount of data is normalized so that the value 1 corresponds to the frame size.  Note that the blue line of the middle graph is similar to the green line of the bottom graph, except that the blue line is flattened out due to packet loss in transmission.  Note that the blue line must be above 1 for each frame to recover the frame, and the closer it is to 1 the closer the bandwidth required to deliver the frame is optimal.  However, when there is more packet loss variance, the amount of data that arrives for the frames (as indicated by more vertical variance in the blue line) is also more variable.  These are the kinds of issues that are accounted for in the overall protocol design.  

The size of the frames are also shown by the red line of the middle graph mapped to the right Y-axis, where the right Y-axis shows the frame size in units of packets.

The delivery latency of each frame is shown by the black line mapped to the left Y-axis in the top graph of Figure~\ref{Slowfig}.  As shown, the delivery latency is close to the minimal possible 20-25 ms as packet loss conditions vary, while at the same time the amount of \liqdata\ sent to deliver the frame is close to optimal.

Figure~\ref{Fastfig} shows the results for a 100 Mbps video stream at 30 fps, where the frame sizes varied in a repeating pattern of one frame of size 1.25 MB followed by 5 frames of size 250 KB each.  This represents an I-frame followed by five B-frames in a repeating pattern.  The packet loss was varied in steps between $0.3\%$, $2\%$, $6\%$, $11\%$, $6\%$ and $0.3\%$ as shown in the red line mapped to the right Y-axis in the bottom graph of Figure~\ref{Fastfig}. 

The results shown in Figure~\ref{Fastfig} are in the same format as those shown in Figure~\ref{Slowfig} described above.  Similar to the 9.6 Mbps video stream shown in Figure~\ref{Slowfig}, the delivery latency for the 100 Mbps video stream shown in Figure~\ref{Fastfig} is close to the minimal possible 25-30 ms as packet loss conditions vary, while at the same time the amount of \liqdata\ sent to deliver the frame is close to optimal. As is evident from the results, the \oursoftware\ is effective is real world use cases and deployments. 

\section{Future work}

For wide-scale deployment, we have developed a tunnel interface for the \oursoftware\ that allows seamless integration with existing applications. The tunnel interface is similar to a VPN tunnel, tunneling all network traffic exchanged between the sender and the receiver through the \oursoftware. We are testing the tunnel interface through field trials and forward deployments and the results look promising -- showing the benefits of \oursoftware\ in real world ultra-low latency use cases.

The two key properties of RaptorQ fountain code of being expandable and interchangeable allow for many interesting use cases. One such use case is enabling \liqdata\ to be transmitted from the sender to receiver over multiple network paths, thus providing both reliability and bandwidth aggregation. Such a service/device can be used to provide better connectivity between enterprise branch offices, home offices etc. and cloud endpoints.  Initial versions of this capability are integrated into the \oursoftware, and performance results will become available in the future.

A key aspect of data delivery is rate control, i.e., the automatic dynamic adjustment of the transmission rate as available bandwidth varies.  Rate control is especially important when dealing with wireless networks where the amount of available bandwidth can suddenly change due to varying signal quality.  Designing and integrating rate control solutions into the \oursolution s described in this work is a crucial next step in this work.

Another key aspect of data delivery is network buffering policies.  It is well known that buffer bloat, i.e., the phenomena where network buffers fill with data that causes long delivery latencies, is an anathema to enabling immersive experiences.  An important future research direction is to understand the interactions between buffering policies and the \oursolution s described in this work, which will lead to improved immersive experience support.

Although the network conditions described in Section~\ref{impaired network section} are based
on conditions observed in a major online cloud gaming provider's current deployment, it would 
still be very useful to provide results based on an actual deployment.  This will be the focus
of future work as these solutions are more widely deployed.

\section{Conclusions}
The results presented in this paper demonstrate a data delivery solution designed from the ground up on the RaptorQ fountain code that provides a practical and compelling approach to enabling immersive experiences over networks that experience fluctuating network conditions.


\input{main.bbl}

\end{document}